\documentstyle[editedvolume]{crckapb}

\begin{opening}

\title{THE EUROPEAN LARGE AREA ISO SURVEY:\protect\\
       ELAIS
}
\author{S.J. OLIVER}
\institute{Imperial College of Science Technology and Medicine\\
	   Astrophysics Group\\
	   Blackett Laboratory\\
	   Prince Consort Rd.\\
	   London\\
	   SW2 1BZ\\
	   {\em s.oliver@ic.ac.uk}}

\end{opening}

\def\deg{\hbox{$^\circ$}}
\def\sun{\hbox{$\odot$}}

\def\lesssim{\mathrel{\hbox{\rlap{\hbox{\lower4pt\hbox{$\sim$}}}\hbox{$<$}}}}
\def\gtrsim{\mathrel{\hbox{\rlap{\hbox{\lower4pt\hbox{$\sim$}}}\hbox{$>$}}}}

\def\arcmin{\hbox{$^\prime$}}
\def\arcsec{\hbox{$^{\prime\prime}$}}

\def\micron{\hbox{$\mu$m\ }}
 
\begin{document}

\begin{abstract}
\footnote{To appear in ``Cold Gas at High Redshift'' Eds:
            Malcolm Bremer, Huub Rottgering, Paul van der Werf
            and Chris Carilli. Pubs: Kluwer}

I describe a European collaborative project to survey $\sim 20$ square
degrees of the sky at 15\micron and 90\micron with ISO.  This is the
largest open time project being undertaken by ISO.  The depth and
areal coverage were designed to complement the various Guaranteed Time
surveys.  The main science thrust is to explore star formation in
galaxies to a much higher redshift than was probed by IRAS. We expect
to detect around 8000 extra-galactic objects and a similar number of
Galactic sources.  The maps and source catalogues will represent a
major legacy from ISO, inspiring follow up work for many years to
come.

\end{abstract}

\section{Introduction}

The Infrared Space Observatory (ISO) will be the
only major infrared mission for the next decade.
Although the satellite was principally designed
as an observatory the case for devoting a substantial
amount of the mission time to surveys was overwhelming.

The Infrared Astronomical Satellite (IRAS) had enormous success
arising principally from its survey products (particularly the Point
Source Catalog and the Faint Source Catalog).  Perhaps most
significant was the discovery of a whole new class of objects with
enormously high far infrared luminosity [notably F10214+4724
\cite{rr91} and P09104+4109 \cite{klein88}].  As well as discovering
new objects, IRAS demonstrated the benefit of selecting objects in the
far infrared.  This wave-band is not sensitive to dust obscuration
which biases optically selected samples.  The emission arises from
thermally heated dust and thus complements studies of emission
directly from star-light, gas, or AGN engines.

The sensitivity of ISO is orders of magnitude better than IRAS.
Using it as a survey instrument will thus allow us to explore
IRAS-like populations to higher redshift and possibly unveil
new classes of objects or unexpected phenomena.

This paper outlines the open time survey which is a collaborative
venture between fifteen European Institutes; the PI being
M. Rowan-Robinson, Co-Is being: C. Cesarsky, L. Danese,
A. Franceschini, R. Genzel, \linebreak A. Lawrence, D. Lemke, R. McMahon,
G. Miley, S. Oliver, J-L. Puget and B. Rocca-Volmerange.  Many other
people are also heavily involved.

\section{Science Goals}

While it is impossible to predict all the scientific benefits of such
a large project, I outline some of the key issues that we hope
to address.  A major theme is the detection of high redshift galaxies.

\subsection{Epoch of Galaxy Formation}

The search for galaxies at high redshift to uncover the formation
epoch is one of the holy grails of cosmology.  The failure to detect
high redshift objects in optical surveys, particularly using
Ly$\alpha$, has two competing explanations.  The first is that early
galaxies contain a large dust component which obscures the optical
emission.  E.g.  if elliptical galaxies underwent a massive burst of
star-formation between $2<z<5$, they would be observable in the far
infrared since massive stars produce both dust and the UV to heat it,
and may look like F10214 \cite{de92}.  Alternatively galaxies may have
been formed by the assembly of constituents which are individually too
faint to detect.  This survey will provide a powerful discrimination
between these two hypotheses, since we would detect optically obscured
galaxies but not low luminosity proto-galaxies.

\subsection{Star Formation in Spiral Galaxies at High Redshift}

The main extra-galactic population detected by IRAS was galaxies with
high rates of star formation. Their far infrared emission arises from
dust heated by young stellar populations.  These objects are now known
to evolve with a strength comparable to AGN \cite{so95}.  The distance to
which these objects were visible by IRAS was, however, insufficient to
determine the nature of their evolution.  The sensitivity of ISO will
allow us to detect these objects at much higher redshifts and thus
obtain greater understanding of the cosmological evolution of star
formation.

\subsection{Ultra and Hyper-luminous I-R Galaxies at High $z$}

IRAS uncovered a population with enormous far infrared luminosities,
$L_{\rm FIR} $ $> 10^{12}L\sun$.  This far infrared emission
represents the bulk of the bolometric luminosity of these objects
which is comparable to that of AGN.  The local space density of these
objects, however, exceeds that of optically selected AGN, implying this
population is a more energetically significant component of the
Universe.  For these reasons this population has been carefully
studied.  The energy source in these objects is still disputed.
While most of these objects appear to have an AGN it is argued that
star formation could provide most of the energy.  Interestingly, most
of these objects appear to be in interacting or merging systems,
suggesting a triggering mechanism.  Exploration of these objects at
higher redshift will have particular significance for models of
AGN/galaxy evolution.

\subsection{Emission from Dusty Tori around AGN}

Unified models of AGN suggest that the central engine is surrounded by
a dusty torus.  Optical properties are then dependent on the
inclination angle of this torus.  The far infrared emission from the
torus will be less sensitive to the viewing angle.  Thus a far
infrared selected sample of AGN will be more uniform than an optically
selected sample and the far infrared properties of these will place
important constraints on unification schemes.  AGN are known to be
strongly evolving and this sample will tell us about the evolution of
the tori.  Also, we will be able to detect dust emission from tori in
`face-on' AGN which would  not be detected in the optical.

\subsection{Dust in Normal Galaxies to Cosmological Distances}

Faint optical redshift surveys find surprisingly few galaxies beyond
$z=0.5$.  One possible explanation for this is a dust fraction that
increases with $z$.  Emission from the cool interstellar `cirrus' dust
in normal galaxies will be detectable in our survey to much greater
distances than were accessible with IRAS, so we will be able to
examine the dust content to higher $z$.

\subsection{Circumstellar Dust Emission from Galactic Halo Stars}

The deep stellar number counts provided by this survey will be relatively
unaffected by Galactic extinction and may provide, amongst other things,
improved estimates of the halo/disk population ratios.

\subsection{New classes of Galactic and Extra-Galactic Objects}

F10214 was at the limit of IRAS sensitivity and new classes of objects
may well be discovered at the limit of the ISO sensitivity.  The
lensing phenomenon which made F10214 detectable by IRAS may become more
prevalent at fainter fluxes, increasing the proportion of interesting
objects.  Current predictions suggest we would not expect to detect
Galactic Brown Dwarfs but unexpected Galactic objects may be
discovered.

\subsection{Clustering Properties}

The volume of this survey is comparable to that surveyed by the entire
IRAS Point Source Catalog.  The median redshift will be much higher.
We will thus be in a position to examine the evolution of clustering
strength, giving perhaps the most direct test of the gravitational
instability picture of structure formation.

\section{Survey Definition}

As with any time-constrained survey we had to balance factors such as
depth, wavelength and areal coverage.  To complement Guaranteed Time
deep ISO CAM surveys \cite{af95} we decided to sacrifice depth at the
shorter wavelength for increased areal coverage.  This section
describes the rationale behind the choices we made for: wavelengths,
depths and areas.

\subsection{WAVELENGTH AND SENSITIVITIES}

We initially proposed to survey at three wavelengths to give useful
colour information over a long wavelength baseline but were required
by the OTAC to restrict ourselves to two.  At the longer ISO
wavelengths we pick up star forming galaxies.  Consideration of the
SED of these galaxies together with the capabilities of the ISO PHOT
instrument suggested that the optimal sensitivity to these objects
would be obtained using the C100 detector with 90\micron filter.  At
shorter wavelengths ISO is more sensitive to AGN emission.
Consideration of the ISO CAM sensitivities, AGN SEDs and avoidance of
frequencies in atmospheric windows lead us to select the CAM LW-3
filter centred at 15\micron.

The limited resolution but high sensitivity of ISO at long wavelengths
means that the Galactic Cirrus confusion limit is reached with very
short integration times.  This confusion limit thus defined our PHOT
integration.  We decided to use a similar total observation time for
both instruments.  Table \ref{aot} summarises the two observing modes
used.

\begin{table}
\caption{Survey parameters for a single raster}\label{aot}
\begin{tabular}{lcc}\\
Instrument                &    CAM          &    PHOT           \\
\\
Filter                    &  LW-3           &  90               \\
$\lambda_0/\micron$       & $15\pm3$        & $95.1\pm26$       \\
Detector       	          &  LW-l Si(+Ga)   &  C100 Ge:Ga       \\
AOT                       &  CAM01          &  PHOT22           \\
Pixel Size                &   5.6\arcsec    &   43.5\arcsec     \\   
Pixels/frame              & $32\times 32$   & $ 3\times 3 $     \\
Frame size                & 180\arcsec      &  135\arcsec       \\
$\delta x, \delta y$      & 90\arcsec,180\arcsec& 130\arcsec,130\arcsec \\
Raster Points             & $28\times14$    & $20\times20$      \\
Raster size               & $(42\arcmin)^2$     & $(43.3\arcmin)^2$     \\
5$\sigma$ Sensitivity     & 1.7 mJy            & 15 mJy \\
\\
\end{tabular}

\end{table}

\subsection{AREAS}

The allocated observing time allowed 37 rasters as described above.
The choice of where to distribute these on the sky was governed by a
number of factors.  Firstly we decided not to group these all in a
single contiguous region of the sky.  Had we done so we may have had
difficulty distinguishing evolutionary effects from local large scale
structures.  Distributing the survey areas across the sky also has
advantages for follow up work.  Cirrus confusion is a particular
problem, so we selected regions with low IRAS 100\micron intensities
($I_{100}<1.5$MJy/sr), using the maps of \cite{rr91b}.  To avoid
conflict with other ISO observations we further restricted ourselves
to regions of high visibility over the mission lifetime ($>25$\%).  To
avoid unnecessarily high Zodiacal backgrounds we only selected regions
with high Ecliptic latitudes ($|\beta|>40^\circ$).  Finally it was
essential to avoid saturation of the CAM detectors so we had to avoid
any bright IRAS 12\micron sources.  These requirements led us to
selecting the four areas detailed in Table \ref{areas}.  A further 6
areas were selected as being of particular interest to warrant a
single small ($24\arcmin\times 24\arcmin$) raster.  These were chosen
either because of existing survey data or because the field contained
a high redshift object and were thus more likely to contain high
redshift ISO sources.  These 6 regions are also described in Table
\ref{areas}.

\begin{table}
\caption{Summary of Areas.  The first four areas comprise the
main survey made up from $43\arcmin\times43\arcmin$ rasters.
One raster in N3 will be repeated.  The final 6 areas
are single smaller rasters $24\arcmin\times24\arcmin$}\label{areas}
\begin{tabular}{llrrccc}
Area & 
Rasters &
\multicolumn{2}{c}{Nominal Coordinates} & 
 $\langle I_{100}\rangle$ &
 Visibility &
 $\beta$ \\
 & & \multicolumn{2}{c}{J2000} 
 & \begin{small}$/{\rm MJy sr}^{-1}$ \end{small}& /\% & \\
\\
N1 & $3\times 3$ &$16^h08^m44^s$&$+56\deg26\arcmin30\arcsec$& 
 1.2 & 98.0 & 73 \\
N2 & $4\times2-1$ &$16^h39^m34^s$&$+41\deg15\arcmin34\arcsec$&
 1.1 & 58.7 & 62 \\
N3 &$ 3\times3$ &$14^h28^m26^s$&$+32\deg25\arcmin13\arcsec$&
 0.9 & 26.9 & 45 \\
S1 &$4\times3$ &$00^h38^m24^s$&$-43\deg32\arcmin02\arcsec$& 
 1.1 & 32.4 &-43 \\
\\
Lock. 3  & 1 &$13^h34^m36^s$&$+37\deg54\arcmin36\arcsec $ & 
 0.9 & 17.3 & 44 \\ 

Sculptor   & 1 & $00^h22^m48^s$&$-30\deg06\arcmin30\arcsec $&
 1.3 & 27.5 & -30 \\  

TX1436 & 1 & $14^h36^m43^s$&$+15\deg44\arcmin13\arcsec $&
 1.7 & 22.2 & 29 \\
4C24.28    & 1 & $13^h 48^m 15^s $&$ +24\deg 15\arcmin 50\arcsec $&
 1.4 & 16.8 & 33 \\
VLA 8 & 1 & $ 17^h 14^m 14^s $&$+50\deg 15\arcmin 24\arcsec $&
 2.0 &  99.8  &   73 \\
Phoenix  & 1 & $ 01^h 13^m 13^s $&$ -45\deg 14\arcmin 07\arcsec $ &  
 1.4 &  36  &  \\

 \end{tabular}

 \end{table}

\section{Expectations}

IRAS luminosity functions and model SED of star-bursts, AGN and normal
galaxies and F10214 like objects together with simple pure luminosity
evolution models have been used to predict the number of
extra-galactic object we expect to see \cite{cpp95}.  This simple model
predicts 5000 star-bursts (20\% detected in both bands, 30\% $z>1$), 650
AGN (5\% detected in both bands, 23\% $z>1$), 2300 normal galaxies
(30\% detected in both bands) and 4 F10214 like objects.  Models
including a dusty phase in elliptical galaxy formation would predict
higher numbers.  We would also expect of order 10000 stars.

\section{Science Products}

The products we will provide to the community are source catalogues
together with catalogue associations and maps at both ISO wavelengths.
We anticipate these will be available a year after the end of the ISO
mission (i.e. May 1998).  A WWW page will be on line in the near
future \footnote{available now http://artemis.ph.ic.ac.uk/} to keep
the community abreast of the progress of the survey, a link to this
will be found on http://icstar5.ph.ic.ac.uk/


\begin{thebibliography}{}


 

\bibitem[\protect\citeauthoryear{Elbaz {\it et al.}}{1992}]{de92}
Elbaz, D. {\it et al.} (1992) {\it Astr. Astrophys.,}{\bf   Vol.~no.~265},
 pp.~L29--L32

\bibitem[\protect\citeauthoryear{Franceschini {\it et al.}}{1995}]{af95}
Franceschini A., Cesarsky,C Rowan-Robinson, M., (1995) 
In `Near-IR Sky Survey'
San Maniato (Pisa){\it Memorie della Societa Astronomica Italiana} (in press)

\bibitem[\protect\citeauthoryear{Kleinmann {\it et al.}}{1988}]{klein88}
Kleinmann, S.G. {\it et al.} (1988) {\it Astophys. J.} {\bf Vol.~no.~328},
pp.~161--169


\bibitem[\protect\citeauthoryear{Oliver {\it el al.}}{1995}]{so95}
 Oliver, S., {\it et al.}  (1995), In Wide-Field Spectroscopy and the
 Distant Universe, {\it Maddox, S.J., Aragon-Salamanca, A. eds,
 Proceedings of the 35th Herstmonceux Conference, World Scientific.}
 p. 264

\bibitem[\protect\citeauthoryear{Pearson \& Rowan-Robinson}{1995}]{cpp95}
Pearson, C., Rowan-Robinson, M. {\it et al.} (1995) {\it Mon. Not. R. Astro. Soc.,} 
(in press)

\bibitem[\protect\citeauthoryear{Rowan-Robinson {\it et al.}}{1991}]{rr91}
Rowan-Robinson, M. {\it et al.} (1991)  {\it Nature},  {\bf Vol.~no.~351}, 
pp.~719--721

\bibitem[\protect\citeauthoryear{Rowan-Robinson {\it et al.}}{1991b}]{rr91b}
Rowan-Robinson, M. {\it et al.} (1991) {\it Mon. Not. R. Astro. Soc.,} {\bf
Vol.~no.~249}, pp.~729-741



\end{thebibliography}
\end{document}